\documentclass[graybox]{svmult}

\usepackage{mathptmx}       
\usepackage{helvet}         
\usepackage{courier}        
\usepackage{type1cm}        

\usepackage{makeidx}         
\usepackage{graphicx}        
\usepackage{multicol}        
\usepackage[bottom]{footmisc}

\usepackage{setspace}
\usepackage{amssymb,amsmath}
\usepackage{subfigure}
\makeindex             

\begin{document}

\title*{Evolution of Zipf's Law for Indian Urban Agglomerations  vis-\`{a}-vis Chinese Urban Agglomerations}
\titlerunning{Evolution of Zipf's Law}

\author{
Kausik Gangopadhyay and Banasri Basu
}
\institute{Kausik Gangopadhyay \at Indian Institute of Management Kozhikode, IIMK P.O., Kunnamangalam, Kerala-673570, India. \\ \email{kausik@iimk.ac.in}
\and Banasri Basu \at Physics and Applied Mathematics Unit, Indian Statistical Institute,
203, B. T. Road, Kolkata-700108, India. \\ \email{banasri@isical.ac.in}}
%
%
\maketitle

\abstract{\small{We investigate into the rank-size distributions of urban agglomerations for
India between 1981 to 2011. The incidence of a power law tail is prominent. A relevant question
persists regarding the evolution of the power tail coefficient. We have developed a methodology to
meaningfully track the power law coefficient over time, when a country experience population growth.
A relevant dynamic law, Gibrat's law, is empirically tested in this connection. We argue that these
empirical findings for India goes in contrast with the findings in case of China, another country
with population growth but monolithic political system.}}

\section{Introduction}
It is the job of a scientist to find a mathematical rigour in a natural system, which is apparently anomalous to a casual observer. When a scientist does this in the physical world, the laws discovered are called physical laws. On the other hand, the human society and the institutions created by human beings seem somewhat vulnerable for such laws being maintained. Since human beings have their own desires and wishes, human institutions are often kept outside the purview of physical laws. Nevertheless, if we find some physical laws being observed in the context of human society, not only it will widen the scope of physical laws, but  also usher a novel dimension in the study of social sciences.

We discuss the case of Indian city size distribution and Zipf's law~\cite{1}(alternatively known as Pareto distribution or simply power law) in this backdrop.
Zipf's law, named after linguist George Kingsley Zipf, is a simple empirical law which is often successful in describing the distribution of populations for
various cities in a nation. Zipf noted that the second most common word in the English language (`of') appears at approximately half the rate of the most
common word (`the'). The third most common word (`to') appears at approximately one third the rate of the most common word~\cite{1}. The fact that the law has
been observed in many other spheres makes it even more mysterious. We can set many examples in this context: word usage in human language~\cite{2}, size distribution of islands~\cite{3}, websurfing~\cite{4}, the distribution of wealth and income in many countries~\cite{5,6}, and the size distribution of lunar craters~\cite{7}. The examples also include forest fires~\cite{8}, solar flares~\cite{9}, and football goal distribution~\cite{10}. Recently in a quantitative analysis~\cite{11} of extensive chess databases it is shown that the pooled distribution of all opening weights follows Zipf's law with universal exponent.

In the context of urban economics and regional science, ``Zipf's law'' is synonymous to a remarkable regularity in the distribution of city sizes all over the world. It is also known as the ``Rank-Size Distribution''. This says that the population of a city is inversely proportional to the city's rank among all cities. This could be interpreted in multiple ways.  Let us take a cut-off, say a population of fourteen million. Indeed according to the 2011 Census, there are three Indian metropolises over the population of fourteen million, Greater Mumbai (18,414,288), Delhi (16,314,838) and Kolkata (14,112,536). If we consider another cut-off, which is just half of the previous cut-off, there ought to be double the number of cities over the new cut-off compared to the previous cut-off. We verify that there are exactly six cities with a population of more than seven million - the other three cities being Chennai (8,696,010), Bangalore (8,499,399) and Hyderabad (7,749,334).  If one calculates the natural logarithm of the rank and of the city size (measured in terms of the number of people) and plot the resulting data in a diagram, a remarkable log-linear pattern is obtained, this is the Rank-Size Distribution. If the slope of the line equals minus 1, (as is for example approximately the case for the USA, India, and France,) the relationship is known as Zipf's Law. Zipf's law has repeatedly been shown to hold in the top tails of city size distribution across different countries and periods~\cite{14,15,16,17,18,19}.

Of course, Zipf's law is really not a law at all. It's merely a simple mathematical model that appears to describe some human behavior.  Even more amazingly, Zipf's law has apparently held for at least 100 years. Given the different social conditions from country to country, the different patterns of migration a century ago and many other variables that you'd think would make a difference; the generality of Zipf's law is astonishing. Keep in mind that this pattern emerged on its own, that is, it is ``self-organised''. No city planner imposes it, and no citizen conspire to make it happen. Something is enforcing this invisible law, but we are still in the dark about what that something might be. Many inventive theorists working in disciplines ranging from economics to physics have taken a whack at explaining Zipf's law, but no one has completely solved it. Paul Krugman, who has tackled the problem himself, wryly noted~\cite{17} that ``the usual complaint about economic theory is that our models are oversimplified - that they offer excessively neat views of complex, messy reality. In the case of Zipf's law the reverse is true: we have complex, messy models, yet reality is startlingly neat and simple.'' Zipf's law is popular and thrilling because of its mysterious nature despite being simple.  In the complex human decision of choice of a dwelling place, the existence of such a simple relationship being held is a mystery in itself.

The evolution of Zipf's law coefficient have been studied for countries such as Japan~\cite{kuninaka}, USA~\cite{25}. In this article, we investigate the evolution of Zipf's law coefficient in case of India during 1981 to 2011 using Indian Census data. Even though, Zipf's law is a static phenomenon, it is important to investigate into its dynamic evolution over time partly because that enables us with a clearer understanding of the process of growth of urban agglomerations. Also it is pertinent for the reason that we want to relate the evolution of Zipf's law coefficient with another law related to the dynamic process -- Gibrat's law.  More specifically, Gibrat's law postulates that the mean and variance of the growth rate of an urban agglomeration are independent of its size. It has been demonstrated~\cite{8} that Zipf's law is an outcome of  Gibrat's law. We will expound the case of India -- a country with remarkable population growth and contrast the findings to the Chinese experience.

The organization of this paper is as follows. The second section elaborates the empirical analysis. The last section summarizes our results comparing and contrasting them with the empirical findings in case of China. This section uses findings from an earlier work~\cite{26}.

\section{Empirical Analysis}

\subsection{Zipf's Law for Indian Urban Agglomerations}
\begin{figure}[t]
\centering
\includegraphics[scale=0.65]{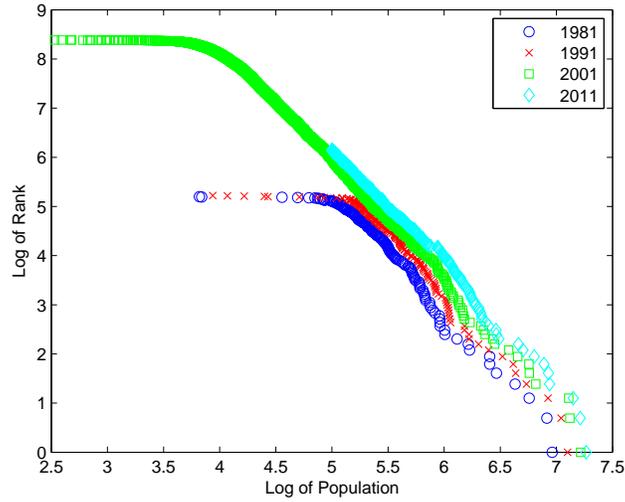}
\caption[]{Rank-size distribution of urban agglomerations in India during 1981-2011.}
\label{fig:Zipfs}
\end{figure}

We gather our data from the Indian Census~\cite{indiancensus}, which is conducted once in a decade. We have used data from four different waves, namely censuses conducted in the years of 1981, 1991, 2001, and 2011. According to the census conducted on the first day of March, 2011, the population of
India stood at 1,210,193,422 persons. The details of Indian census is tabulated in Table \ref{tab:indiancensus}. Figure \ref{fig:Zipfs} shows the rank-size distribution for urban agglomerations of India in 1981, 1991, 2001, and 2011, respectively.

\begin{table}[h]
\begin{tabular}{c|c|c|c}
 Census Year & Total Population &Urban Population  & Rural Population  \\
 & & & \\
 \hline
  2011 & 1,210,193,422 & 377,105,760 & 833,087,662  \\
  &(100)& (31.16)& (68.84)\\
   \hline
  2001 & 1,027,015,247 & 285,354,954 & 741,660,293  \\
  &(100)& (27.78)& (72.22)\\
   \hline
  1991 & 844,324,222 & 217,177,625 & 627,146,597 \\
  &(100)& (25.72)& (74.28)\\
   \hline
  1981 & 683,329,097 & 159,462,547 & 523,866,550  \\
  &(100)& (23.34)& (76.66)\\
   \hline
  \hline
\end{tabular}
\caption{Indian Census during 1981-2011. Figures in parenthesis represent the corresponding figure as percentage of the total population in that particular year.}
\label{tab:indiancensus}
\end{table}

 \begin{figure}[h]
\centering
\includegraphics[scale=0.65]{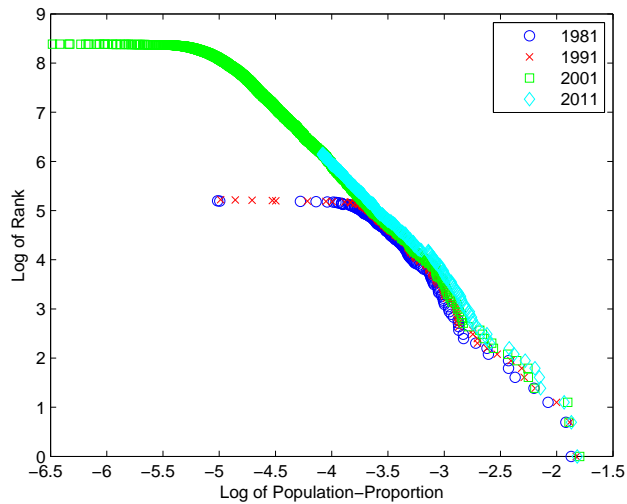}
\caption[]{Rank-population proportion distribution of urban agglomerations in India during 1981-2011.}
\label{fig:Zipfs_proportion}
\end{figure}

The slope of the rank-size line in log scale determines the power law coefficient for any set of urban agglomerations. The estimation for Zipf's law requires a statement on the threshold of the power law region. We can visualise the power law tails for each census year data. However, the starting point of such a tail varies across years. For a country with population growth, it is intuitive that a shift in the threshold of power law region is bound to happen over years. The pertinent issue is: how to handle this question of finding an ``appropriate'' threshold level? We need a measure, which is irrespective of the absolute number of the threshold. This could be materialized, in one way, by considering the population-proportions of urban agglomerations, namely, proportion of population in each urban agglomeration of the total population of India. This will stabilize the threshold value of power law obeying tail over time in a world with considerable population growth. We expound the mathematical form of Zipf's law:
\begin{equation}
\log R(x) = \alpha - \beta \log x,  \hspace{0.2in}\mbox{for all } x > x_0.
\label{eq:zipf}
\end{equation}
The parameter of $\beta$ will be close to unity, under Zipf's law and $x_0$ is the threshold size. For a theoretical abstraction, let us suppose that population of all urban agglomerations grow in equal proportions. Therefore, an urban agglomeration with size $x$ has become one with a  population of $A\cdot x$. The above equation, in that case, boils down to:
\begin{equation}
\log R(x) = \left(\alpha + \beta a\right) - \beta \log Ax, \hspace{0.2in} \mbox{for all } Ax > Ax_0,
\label{eq:zipf_growth}
\end{equation}
where $a = log A$. This Equation (\ref{eq:zipf_growth})  represents the structure of Zipf's law in case a population growth happens. This also implies that the minimum cut-off has to be set upward and the estimated power-law line will have a higher intercept, in such a scenario. Indeed, once the threshold is adjusted, the slope of this line will not change even slightly. Alternatively, if we look into population proportions, we note no change both in the intercept and slope in case of a population growth. Let the population of a country be N. We can calculate the population proportions of urban agglomerations of this country, $\frac{x}{N}$. From a plot of rank-population proportions in the log scale, we can derive a form of Equation (\ref{eq:zipf}). This equation remains unchanged in case the populations of all urban agglomerations and consequently that of the entire country grow by a factor of $A$:
\begin{equation}
\begin{array}{c}
\log R(x) = \left(\alpha - \beta N\right) - \beta \log \frac{x}{N},  \hspace{0.2in}\mbox{for all } x > x_0. \\
\log R(x) = \left(\alpha - \beta N\right) - \beta \log \frac{Ax}{AN},  \hspace{0.2in}\mbox{for all } Ax > Ax_0.
\end{array}
\label{eq:zipf_pop_proportion}
\end{equation}
 We illustrate our case in Figure \ref{fig:Zipfs_proportion} with plots of ranks of urban agglomerations against population-proportions. The curves coincide on one another, mostly. This shows the stability of Zipf's law with respect to relative population over time.
 \begin{figure}[h]
\centering
\includegraphics[scale=0.7]{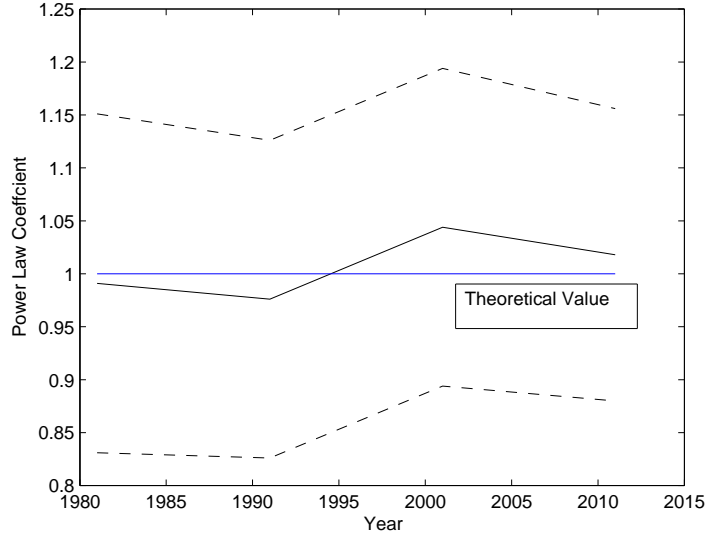}
\caption[]{Evolution of Zipf's law coefficient for India during 1981-2011.}
\label{fig:coefficient}
\end{figure}
 We undertake this strategy of dealing with population growth. The issue boils down to the choice of threshold for urban agglomerations to consider under power law in 1981. Once that decision has been made, we will adjust that threshold for subsequent years by multiplying the initial value by the factor of population growth.  Our focus is to study evolution of Zipf's law coefficient over time. Since we are interested in the evolution of power law tail, the initial value of the power law coefficient hardly matters. We set the threshold for 1981 in such a manner so that the slope of the rank-size plot (in log scale) is as close as possible to unity. This is the value postulated by Zipf's law.

The basic way to estimate the parameters of Pareto distribution is called the ``linear fit method''.  Under this method, we regress the log of rank of an urban agglomeration on the log of its population. The coefficient of the regression line yields the estimate for the exponent of power law. Though widely used, this method produces a biased estimate of the power law exponent \cite{23}.The alternative approach lies in estimation of the Maximum Likelihood Estimator (MLE). This is also known as the ``hill method'' in the econophysics literature. For a sample consisting of a finite number of data points, we can calculate the probability of observing this sample, in entirety, by employing the probability density function and cumulative density function. Given a particular process, this probability is a function of parameters inherent in the particular probability distribution used in the calculation of the sample probability, commonly known as the likelihood of that sample. We maximize this likelihood with respect to the distribution parameters. The set of  parameter-values, for which the likelihood is maximized, is collectively known as the maximum likelihood estimate of the parameters. We have rendered the analytical expressions of the maximum likelihood estimate for the Pareto distribution in our earlier work~\cite{20}.
\begin{figure}[h]
\centering
\subfigure[Scatter Plot for Growth Rate of Urban Agglomerations against Size: 1991-2001]{
\includegraphics[scale=0.38]{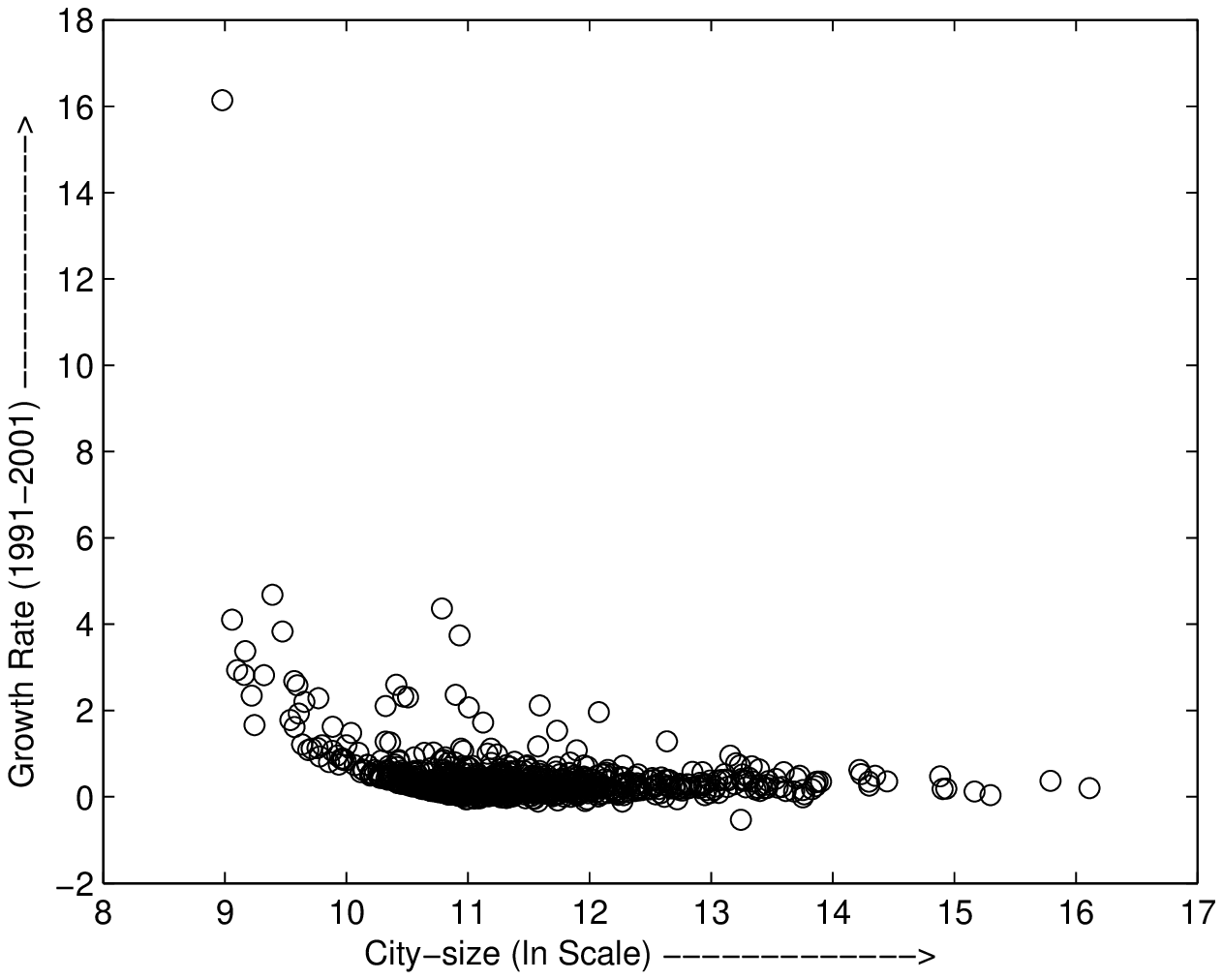}
\label{fig:case-1}
}
\subfigure[Scatter Plot for Growth Rate of Urban Agglomerations against Size: 2001-2011]{
\includegraphics[scale=0.38]{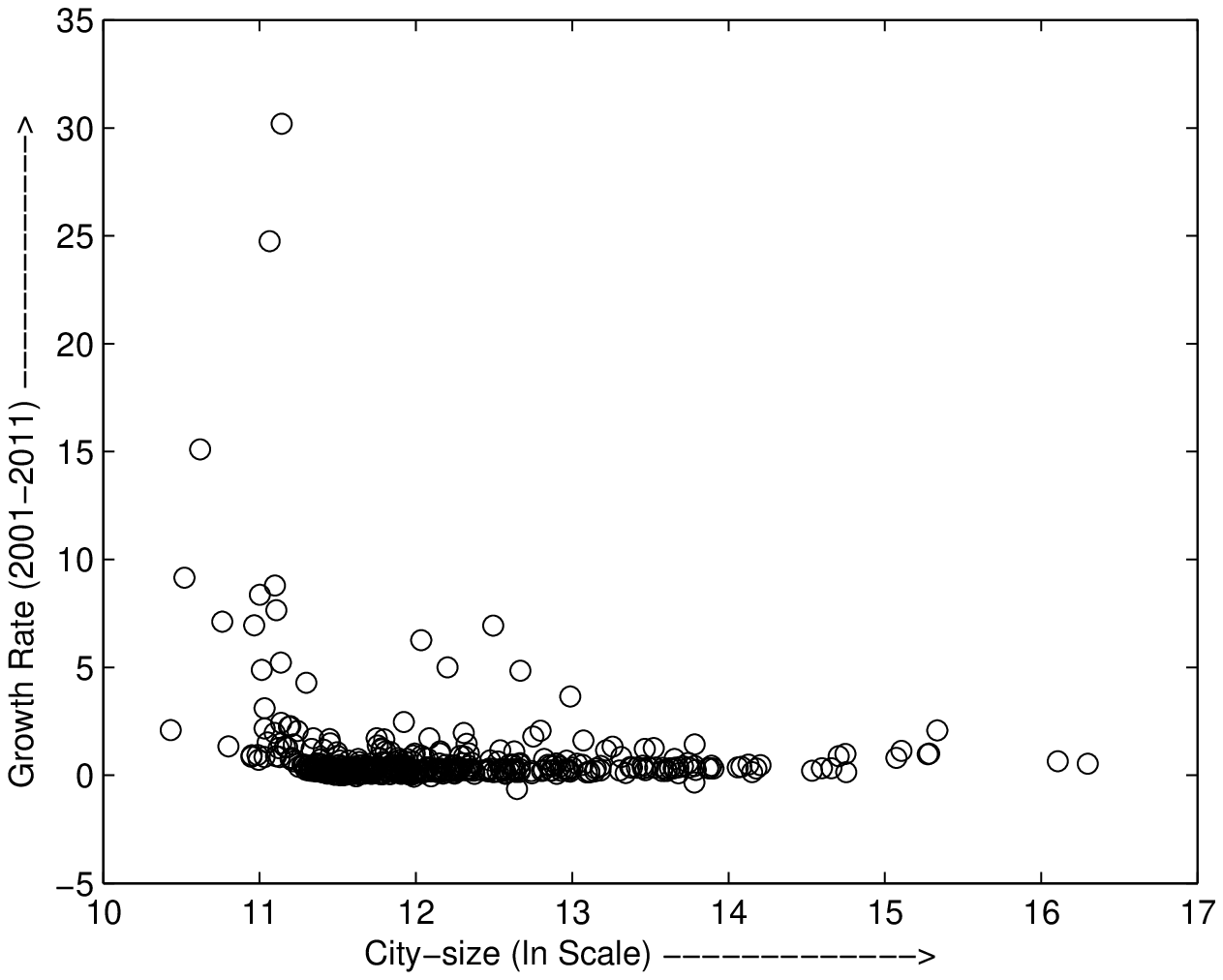}
\label{fig:case-2}
}
\subfigure[Mean Growth Rate of Urban Agglomerations against Size: 1991-2001]{
\includegraphics[scale=0.38]{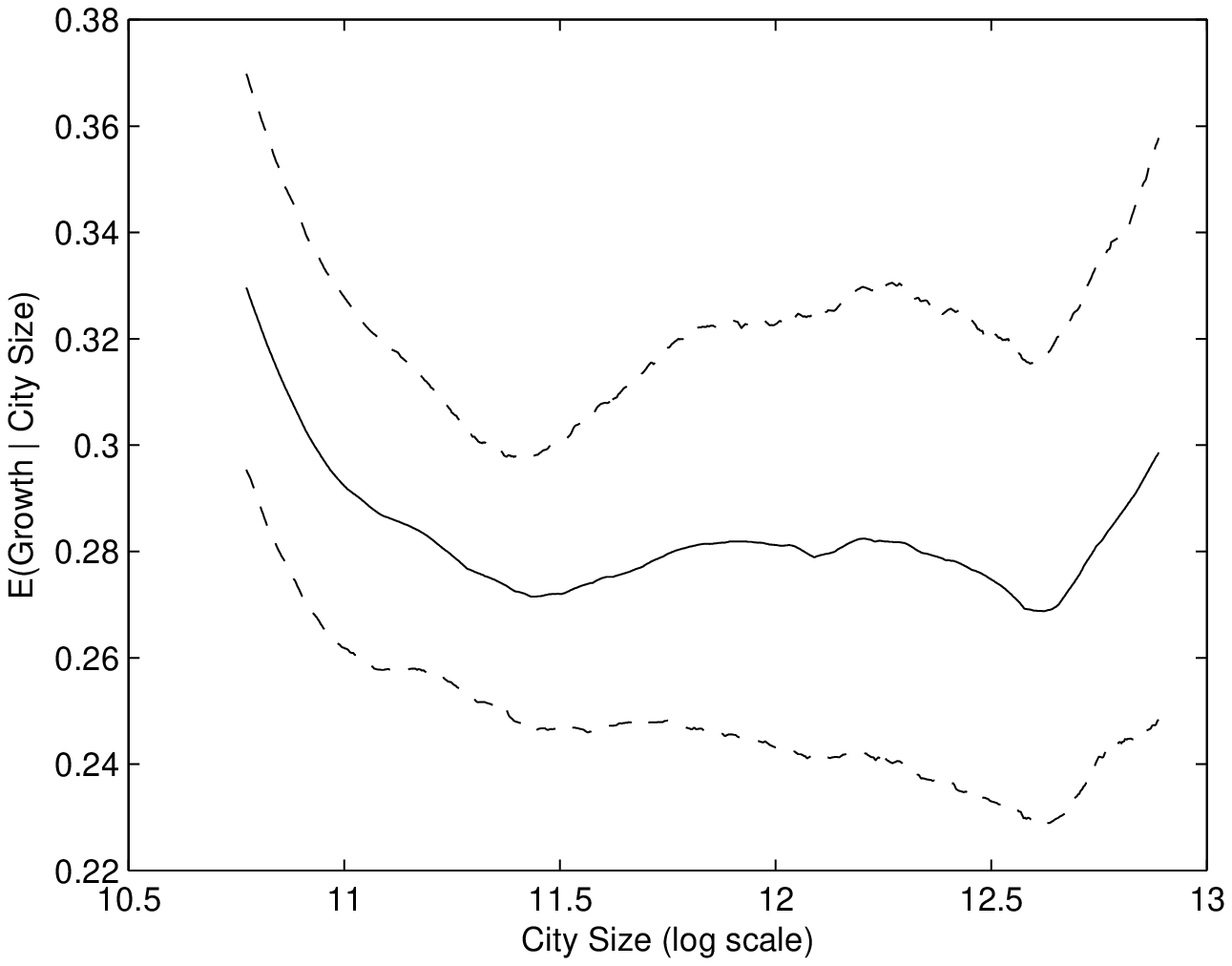}
\label{fig:case1}
}
\subfigure[Mean Growth Rate of Urban Agglomerations against Size: 2001-2011]{
\includegraphics[scale=0.38]{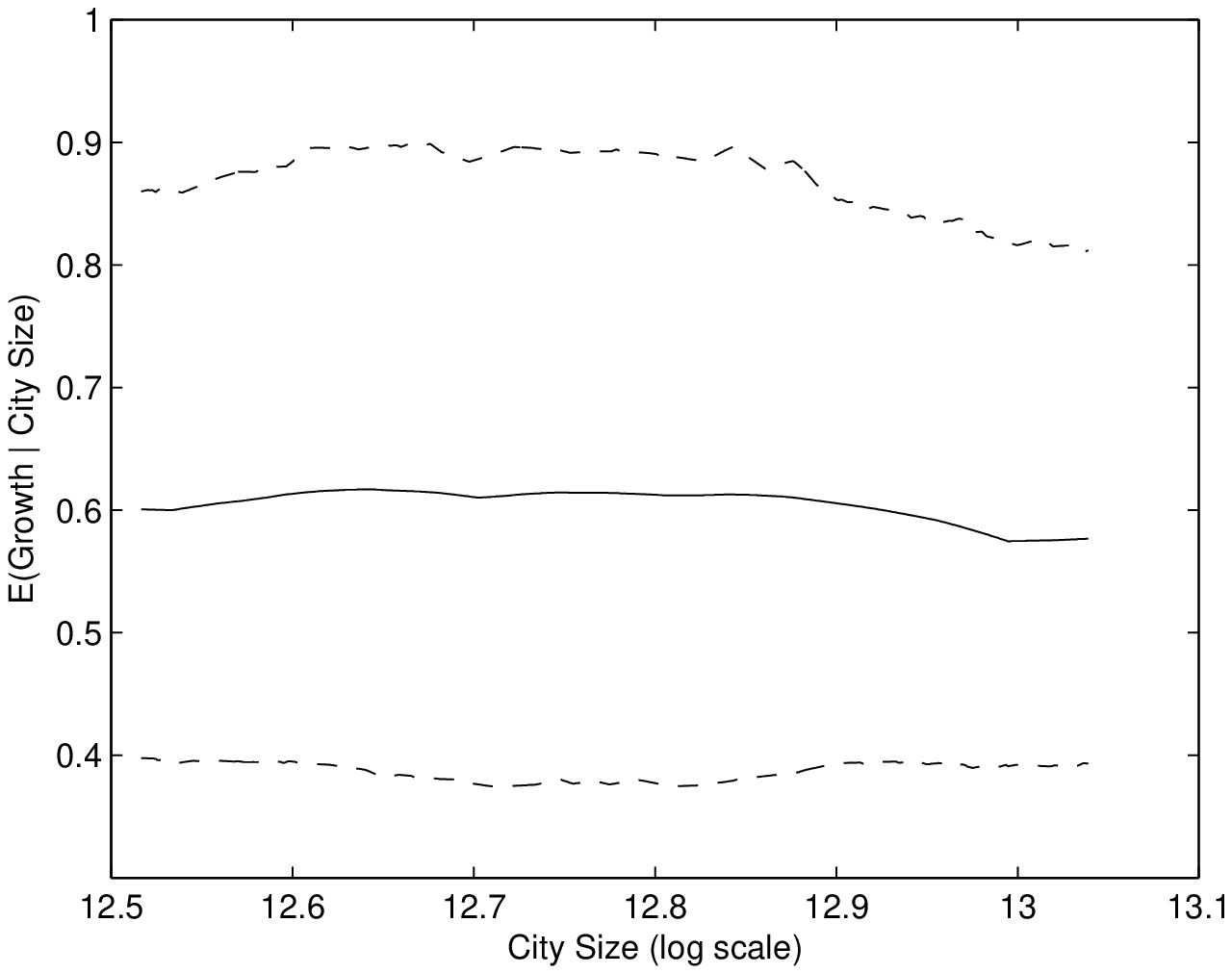}
\label{fig:case2}
}
\subfigure[Growth Rate Variance of Urban Agglomerations against Size: 1991-2001]{
\includegraphics[scale=0.38]{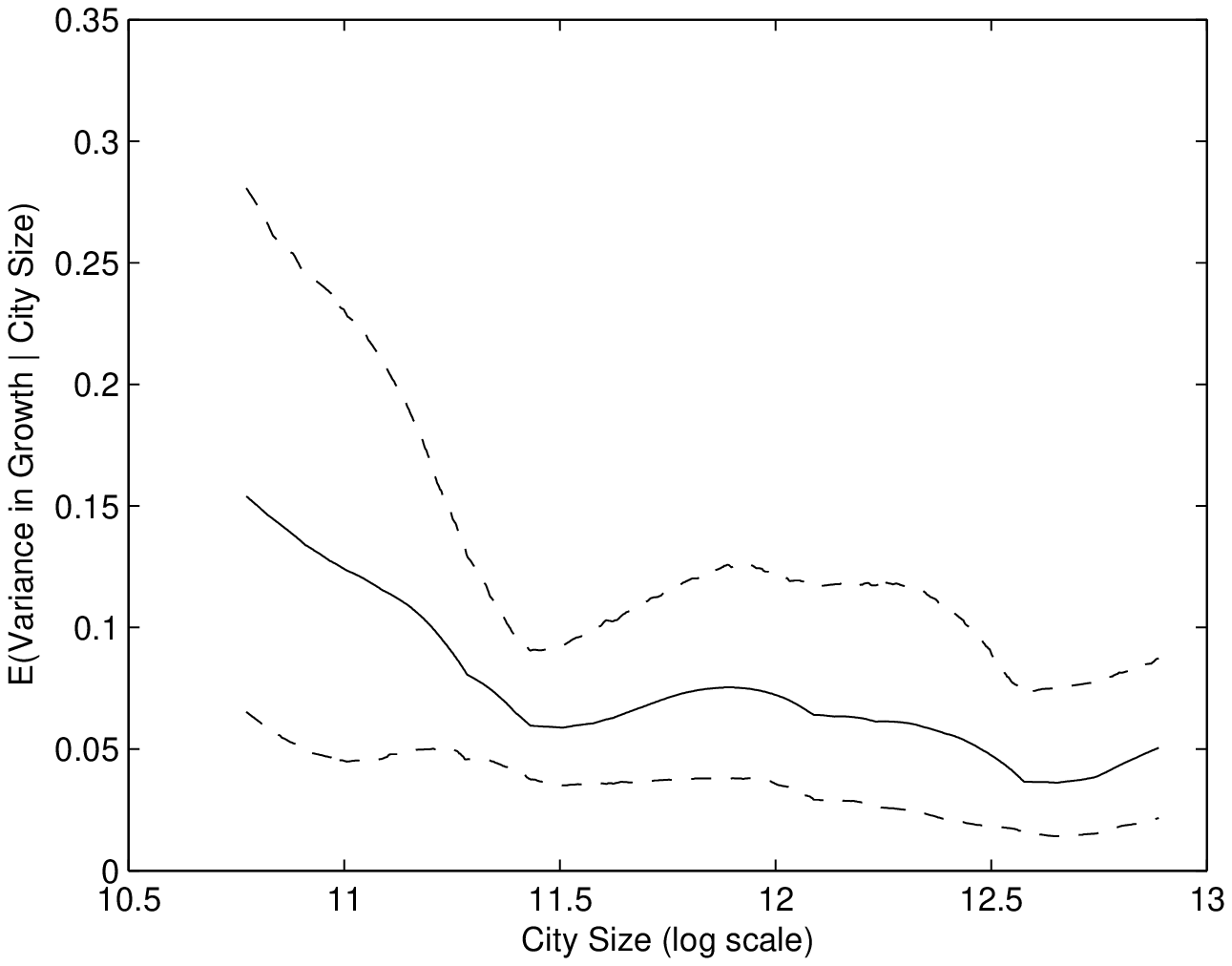}
\label{fig:case3}
}
\subfigure[Growth Rate Variance of Urban Agglomerations against Size: 2001-2011]{
\includegraphics[scale=0.38]{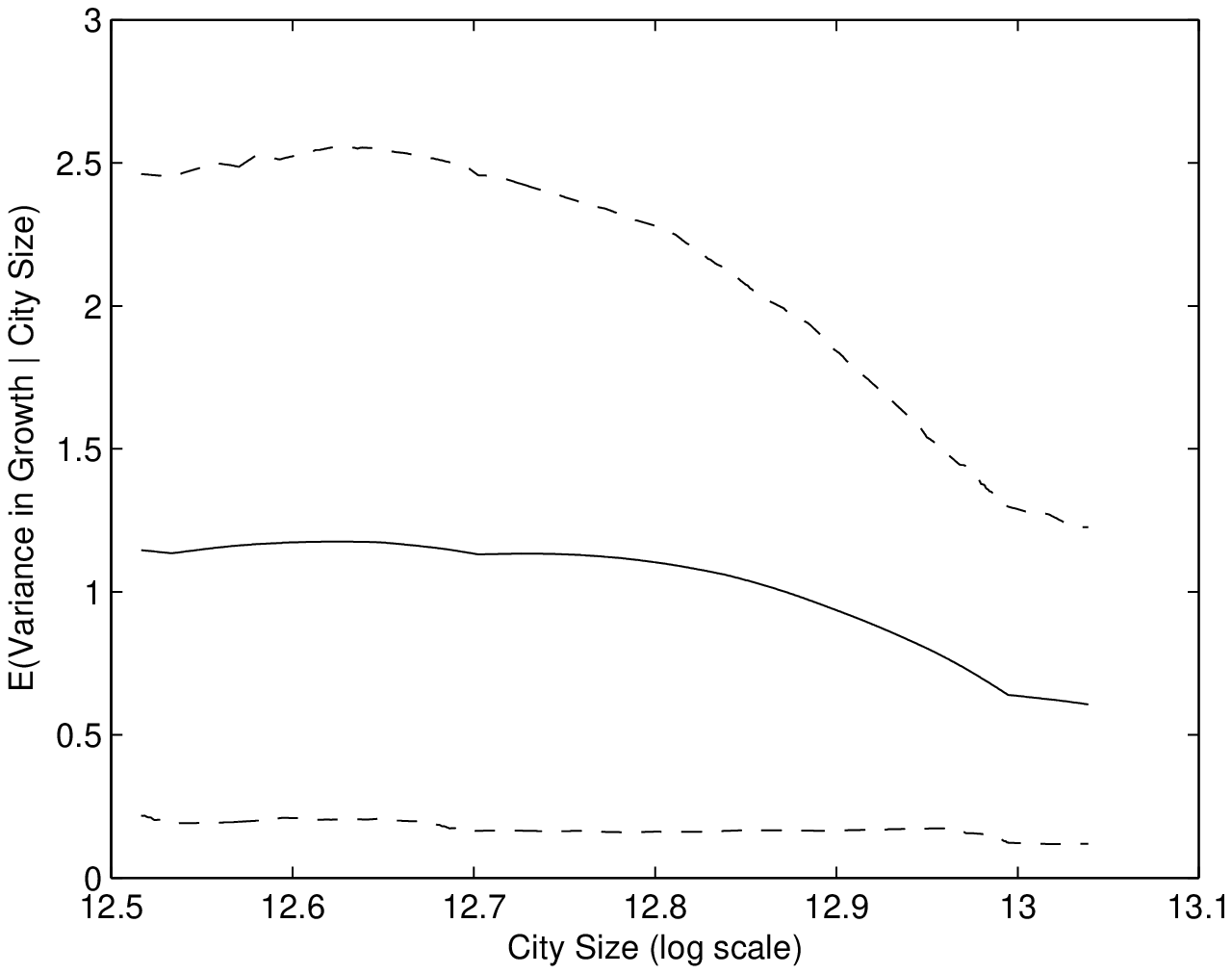}
\label{fig:case4}
}
\caption[]{The growth rates of Indian urban agglomerations have been plotted against their sizes. The scatter plot, mean growth rate (calculated through  Nadaraya-Watson estimate) and the variance of growth rates have been plotted against size of urban agglomerations. Panels (a), (c) and (e) illustrates plots for 1991--2001; whereas panels (b), (d), and (f) are for 2001--2011.}
\label{fig:gibrat_India}
\end{figure}

\begin{table}[h]
\begin{tabular}{c|c|c|c}
 Census Year & Minimum Size &Linear Fit Estimate & Maximum Likelihood Estimate  \\
 \hline
  2011 & 212,523 & 0.935 & 1.018\\
  && (0.007) & (0.069) \\
   \hline
  2001 & 180,355& 0.921 & 1.044\\
  && (0.007) & (0.075)\\
   \hline
  1991 & 148,272 & 0.899 & 0.976 \\
  && (0.008) & (0.075) \\
   \hline
  1981 & 120,000 &   0.889 & 0.991\\
  &&    (0.009) & (0.080)\\
   \hline
  \hline
\end{tabular}
\caption{Zipf's Law Verified for Indian Urban Agglomeration. The standard errors of estimates are noted in the parentheses.}
\label{tab:zipfs}
\end{table}

We have computed both estimates using linear fit method and maximum likelihood method. Our estimates are tabulated in Table \ref{tab:zipfs} for 1981 to 2011 for all four census rounds. As discussed, we have chosen a cut-off of 120,000 as the minimum size of urban agglomeration to be included in 1981 sample so that the Zipf's law coefficient -- estimated value 0.991 -- is close to the theoretical prediction of 1.000. Afterwards, we multiply this minimum value by the population growth rate to arrive at the subsequent figures for the years 1991--2011. For example, the growth rate of population was 23.56\% during 1981--1991. Therefore, we add 23.56\% to the threshold value of 120,000 to arrive at the figure of 148,272 (rounded off to whole number). From Figure \ref{fig:coefficient} it is apparent that the coefficient of Zipf's law is remarkably close to the theoretical value of two for all these years. Also, the movement of Zipf's law coefficient is negligible which indicates little change in the process of formation of urban agglomerations over the course of three decades.

\subsection{Empirical Validation of Gibrat's Law for India}

Next pertinent empirical issue is the validity of Gibrat's law in this context. We examine whether the relation
 between mean and variance of growth rate of urban agglomerations depend on size. We consider the population growth rate
 of all available urban agglomerations for the period of 1991--2001 and 2001--2011. The size for these urban agglomerations are
 their populations in the initial year of the considered period, say 1991 and 2001. A non-parametric way to summarize the
 population growth rates is through the Kernel estimates of local mean, which we elaborate hereby. Suppose, the growth rate of
  a city, $g_i$, bears some relation with the size of the city, $S_i$, modeled as:
\[
g_i = m(S_i) + \epsilon_i,
\]
for all $i = 1, 2, ..., n$, $n$ being the total number of urban agglomerations with available data.
 $g_i$ is the growth rate of the $i$th urban agglomeration in a time period and $S_i$ is its size in the initial year
  of the period considered.

  The objective is to find a smooth estimate of local mean of growth rate over size, say $m(S)$. The Kernel estimate does not give rise
  to inaccuracies in the boundary region.  We choose a particular interval, say $[1.2 \cdot \underset{i}{\min}~ S_i,~~ 0.8 \cdot \underset{i}{\max}~S_i ]$ to exclude the effect of the boundaries. We perform a Kernel density regression in the support of $S_i$. The local average smooths around a point $s$, and the
smoothing is done using a kernel, which is a continuous weight function symmetric around $s$. We use a popular kernel, namely Epanechnikov, for  which: $K(x) = \frac{3}{4}\left(1-\psi^2\right) \cdot \mathbf{1}_{|\psi| \leq 1}$.  The function, $m(\cdot)$, does depend on
the size the bandwidth $h$ of a kernel determines the scale of smoothing. The Nadaraya-Watson
estimate \cite{Pagan_Ullah} of $m(\cdot)$  is given by the following expression,
\[
\hat{m}(s) = \frac{n^{-1} \sum_{i=1}^{n} K_h(s - S_i) g_i}{n^{-1} \sum_{i=1}^{n} K_h(s - S_i)}
\]

The means and variances of the growth rate of the urban agglomerations are illustrated in Figure \ref{fig:gibrat_India}. A visual inspection verifies the veracity of Gibrat's law. There is no significant trend -- either in the mean or in the variance -- over size for both time periods. Thereby, it verifies that since Gibrat's law is satisfied in this case, Zipf's law should hold true, dynamically.

\begin{figure}[t]
\centering
\subfigure[Mean Growth Rate of Urban Agglomerations against Size: 1990-2000]{
\includegraphics[scale=0.38]{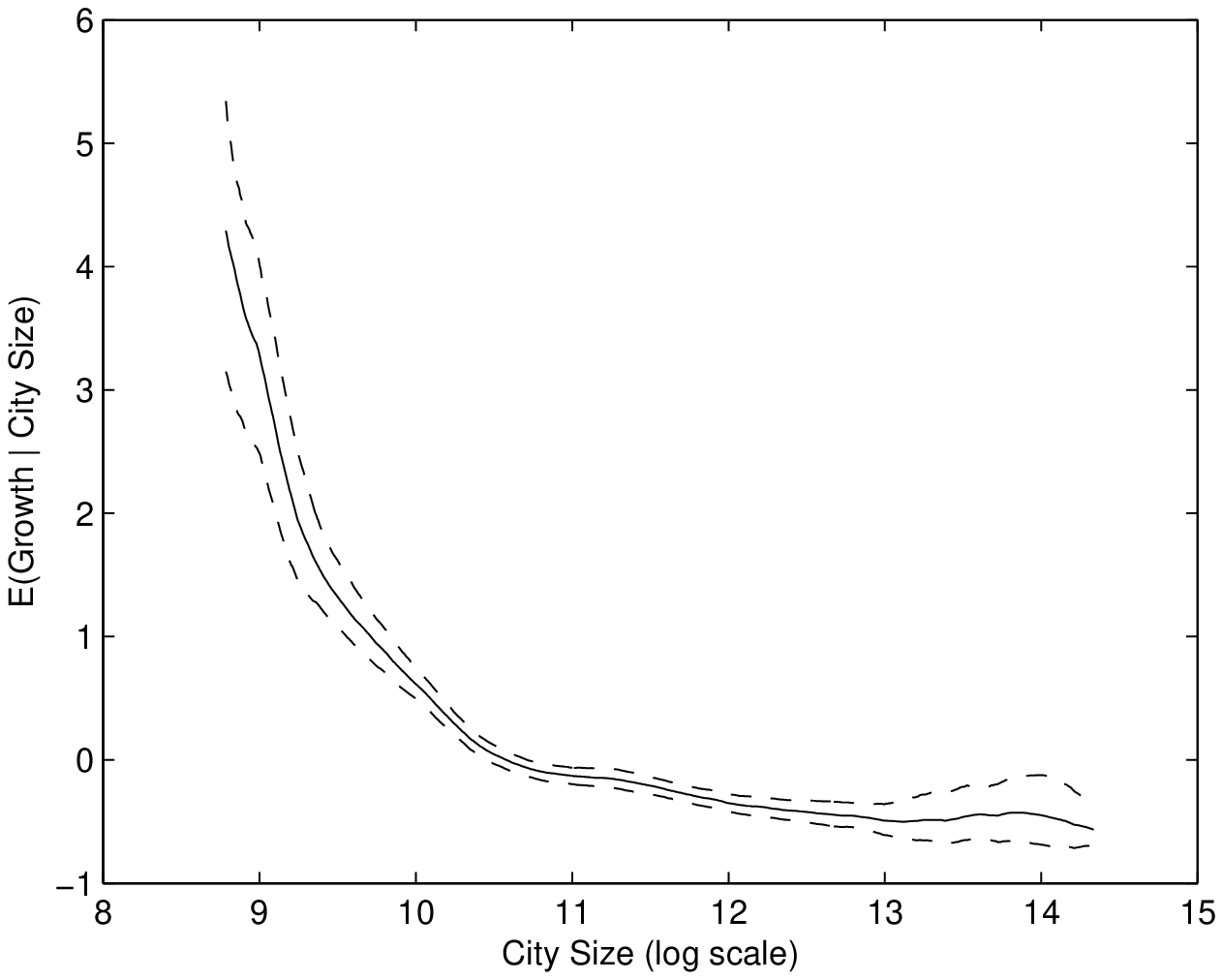}
\label{fig:case1}
}
\subfigure[Growth Rate Variance of Urban Agglomerations against Size: 1990-2000]{
\includegraphics[scale=0.38]{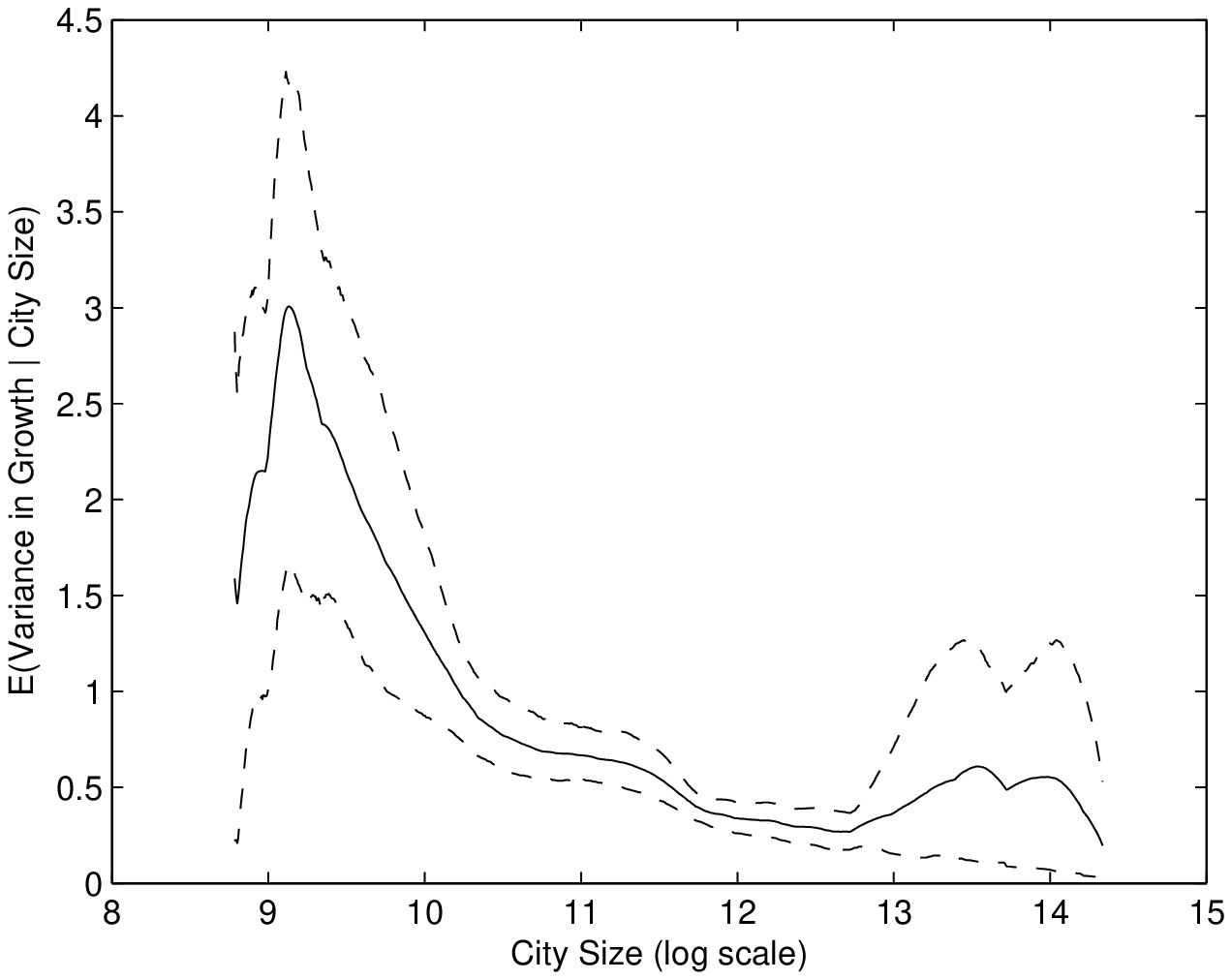}
\label{fig:case2}
}
\caption[]{The mean and variance of growth rates of Chinese urban agglomerations have been plotted against their sizes for 1990--2000. }
\label{fig:gibrat_China}
\end{figure}
\section{Discussion}
India and China are two countries in Asia with a lot of similarities. These are traditionally less urbanized nations and the impetus for urbanization
is fairly recent for both India and China. For India, the urban population augmented from 23.34\% in 1981 to 31.16\% in 2011. Urbanization has taken place in China since the 1980s in an even more rapid scale. The urbanization rate
increases~\cite{29} from 23.01\% in 1984 to 43.90\% by the end of 2006. Definitely, this immense growth opens up a plethora of questions about the morphology in general. For example, cities in special economic zones may have prospered due to favorable government policies unlike their counterparts in non-industrialized zones. First, in this context we demonstrated~\cite{20} that Zipf's law can be equally valid for these countries like urbanized western nations. The subsequent question is about dynamic authenticity of this empirical phenomenon. The Indian experience has already been narrated in Section 2. We sum up our empirical findings: power law coefficient stays near the theoretically prediction over time, 1981--2011, and Gibrat's law is also satisfied. We draw heavily from our past study~\cite{26} regarding the empirical facts on China. The power law coefficient grew between 1990 and 2000. The underlying reason has been detected to a violation in Gibrat's law in case of China. We reproduce the graphs in Figure \ref{fig:gibrat_China}, in which mean and variance of growth rates of the Chinese urban agglomerations during 1990--2000 have been plotted against their sizes. The plots indicate that the large cities grew rather less in China compared to medium and small sized urban agglomerations.

Why  this anomaly between India and China? Is the cause rooted in the policies for formation of Special Economic Zones (SEZs) as surmised before~\cite{26}? Obviously, the growth rate of urban agglomerations is pegged for China at the upper end. The following model of job creation and migration to the newly created SEZs could match the empirical reality in a simulation experiment. The government introduce the feature of Special Economic Zones by giving special privileges to some urban agglomerations. The privileged urban agglomerations are chosen in such a way that they are not from the most populous cities. The other elements of that mathematical model are as follows. The probability of an additional job being created at a location is proportional to the number of already existing jobs at that location. There is a scale parameter, which for certain ranges favours growth of big cities and for other ranges, favours the growth of small cities. A number of new jobs are created in the locations of the SEZs. These new jobs require higher skill levels compared to the previously existing jobs. A worker matched with these jobs leave their old locations of work and move to the new location. Also higher skilled workers are primarily from the top ranking cities. Under these circumstances, a simulation study~\cite{26} demonstrated the augment of power law exponent on account of creation of SEZs.

A verification of the role of SEZ could be experimented with India adopting favourable policies in creating SEZs in India.  The Special Economic Zone Act  was passed by the Government of India in 2005. Subsequent formation of SEZs~\cite{27, 28} could have induced some bias against the growth of urban agglomerations in the upper end. Nevertheless, we fail to notice any significant difference between mean growth rate of urban agglomerations across different sizes in panels (a) and (b) of Figure \ref{fig:gibrat_India}. Therefore, the hypothesis stating the role of SEZs is ineffectual in reconciling this issue when Indian evidence is considered alongwith.

We propose another avenue to resolve this anomaly. It might depend on the government policies on the growth of cities as it is in the case of China.  It has been observed~\cite{29} that the Chinese governmental policies favoured restriction of populous cities to a specific size and expansion of non-populous cities since 1980s. On a more elaborate note, the Chinese government started the urban planning
policy in 1980. This policy and its successor  the Urban Planning Law enacted in 1990 meticulously controls the size of large cities and appropriately develops
medium-sized and small cities. This policy introduces a bias against the growth of large cities and could very well be responsible for non-observance of Gibrat's law. However, that is hardly the case for India. The expansion of rather small cities and restriction of populous cities are not at all promoted in a democratic society like India compared to the policies pursued in a monolithic society like China. Therefore, the distinction may lie in the political system rooted in these two countries.


\begin{thebibliography}{99}
\bibitem{1} G.K. Zipf, Human Behaviour and the principle of Least Effort (Addison-Wesley, Cambridge,MA, 1949)
\bibitem{2} G.K. Zipf, The Psychobiology of Language. Houghton-Mifflin.
\bibitem{3} Y. Sasaki et.al. J. Phys. Soc. Japan.
\bibitem{4} B.A. Huberman et.al. Science 280 (1998) 95.
\bibitem{5} V. M. Yakovenko and J. B. Rosser, Jr., Rev. Mod. Phys. 81 (2009) 1703.
\bibitem{6} A. Chatterjee and B.K. Chakrabarti, Eur. Phys. J. B. 60 (2007) 135.
\bibitem{7} R.B. Baldwin, Astron. J..69 (1964) 377.
\bibitem{8} B.D. Malamud et.al. Science 281 (1998) 1840.
\bibitem{9} G. Boffetra et. al. Phys. Rev. Lett 83 (1999) 4662.
\bibitem{10} L.C. Malacarne, R.S. Mendes, Physica A 286 (2000) 391.
\bibitem{11} B. Blasuis and R. Tonjes,  Phys. Rev. Lett. 103 (2009) 218701.
\bibitem{14} D. Zanette, S.C. Manrubia, Phys. Rev. Lett. 79 (1997) 523.
\bibitem{15} N.J. Moura Jr., M.B. Ribeiro, Physica A 367 (2006) 441.
\bibitem{16}X. Gabaix, Y. Ioannides, Handbook of Regional and Urban Economics 4,V. Enderson and J.-F. Thisseb(eds.) 2004 North Holland, 2341.
\bibitem{17} P. Krugman, The self Organising Economy, (Blackbell Publishers Oxford, UK and Cambridge MA).
\bibitem{18} X. Gabaix, Quarterly J. of Economics 114 (1999) 739.
\bibitem{19} Kwok Tong Soo,  Regional Science and Urban Economics, 35(3) (2005) 239.
\bibitem{kuninaka} H. Kuninaka, and M. Matsushita, J. Phys. Soc. Jpn. 77 (2008) 114801.
\bibitem{25} J. Eeckhout,  American Economic Review 94(5), (2004), 1429.
\bibitem{20} K. Gangopadhyay and B. Basu, Physica A 388 (2009) 2682.
\bibitem{indiancensus} http://www.censusindia.gov.in
\bibitem{23} A. Clauset, C. R. Shalizi,  M. J. Newman, Power law distributions in empirical data, arXiv: 0706.1062v1.
\bibitem{Pagan_Ullah} Adrian Pagan and Aman Ullah, Nonparametric Econometrics, Cambridge University Press, 1999.
\bibitem{26} K. Gangopadhyay and B. Basu, in Econophysics and Economics of Games, Social Choices and Quantitative Techniques (ISBN: 978-88-470-1500-5), 90.
\bibitem{27} P. Nema,  and P. Pokhariyal,  SEZs as Growth Engines - India vs China (October 5, 2008), SSRN Working Paper.
\bibitem{28} A. Sen , Will India recreate China's SEZ magic? The Economic Times, Nov. (2006)
\bibitem{29} Z. Chen, S. Fu, and D. Zhang, Searching for the Parallel Growth of Cities, SSRN Working Paper.



\end{thebibliography}
\end{document}